\documentclass[conference]{IEEEtran}
\usepackage{amssymb}
\usepackage{times,amsmath,epsfig,float}
\usepackage{subfigure}
\usepackage{graphicx}
\begin{document}
%
\title{Minimum BER Power Adjustment and Receiver Design for Distributed Space-Time Coded Cooperative MIMO Relaying Systems}
%
%
%

\author{\IEEEauthorblockN{Tong Peng, Rodrigo C. de Lamare}
\IEEEauthorblockA{Communications Reasearch Group, Department of Electronics\\University of York, York YO10 5DD, UK\\
Email: tp525@ohm.york.ac.uk; rcdl500@ohm.york.ac.uk}
\and
\IEEEauthorblockN{Anke Schmeink}
\IEEEauthorblockA{UMIC Research Centre\\RWTH Aachen University, D-52056 Aachen, Germany\\
Email: schmeink@umic.rwth-aachen.de}}

\maketitle
\IEEEpeerreviewmaketitle

\begin{abstract}
An adaptive joint power allocation (JPA) and linear receiver design
algorithm using the minimum bit error rate (MBER) criterion for a
cooperative Multiple-Input Multiple-Output (MIMO) network is
proposed. The system employs multiple relays with Distributed
Space-Time Coding (DSTC) schemes and an Amplify-and-Forward (AF)
strategy. It is designed according to a joint constrained
optimization algorithm to determine the MBER power allocation
parameters and the receive filter parameters for each transmitted
symbol. The simulation results indicate that the proposed algorithm
obtains performance gains compared to the equal power allocation
systems and the minimum mean square error (MMSE) designs.

\end{abstract}

\section{Introduction}
MIMO communication systems employ multiple collocated antennas at both the source node and the destination node in order to obtain the diversity gain and combat multi-path fading in wireless links. The different methods of STC schemes, which can provide a higher diversity gain and coding gain compared to an uncoded system, are also utilized in MIMO wireless systems for different sizes of the transmitter and different conditions of the channel. Cooperative MIMO systems, which employ multiple relay nodes with antennas between the source node and the destination node as a distributed antenna array, apply distributed diversity gain and provide copies of the transmitted signals to improve the reliability in the wireless communication system \cite{Scaglione}. Among the links between the relay nodes and the destination node, a cooperation strategy, such as Amplify-and-Forward (AF), Decode-and-Forward (DF), and Compress-and-Forward (CF) \cite{J. N. Laneman2004} and various DSTC schemes in \cite{J. N. Laneman2003}, \cite{Yiu S.} and \cite{RC De Lamare} can be employed.

Since the benefits of the cooperative MIMO systems are noticed, extensive studies of the cooperative MIMO networks have been operated \cite{Tang}-\cite{Farhadi G.}. In \cite{P1}, an adaptive joint relay selection and power allocation algorithm based on the MMSE criterion is designed. After each transmission, the optimal relay nodes with minimum MSE will be selected and the power allocation vectors will be fed back to the relay nodes. A joint transmit diversity optimization and relay selection algorithm for the DF cooperating strategy is designed in \cite{P2}. A transmit diversity selection matrix is introduced at each relay node in order to achieve a better MSE performance by deactivating some relay nodes. In \cite{Tang}, an optimal design for the amplify matrix at the relay node using the AF cooperating strategy is derived. The simplest three-node cooperative MIMO network, which contains one source node, one destination node and only one relay node in the system, is employed. The algorithm in \cite{Tang} indicates that the optimal amplify matrix for AF strategy depends on the channel matrices when the direct link between the source node and the destination node is ignored. In \cite{Y. Shi}, a selection algorithm for one relay node among the other neighboring relay nodes is derived. However, with the simplicity of selecting only one relay node to construct the three-node cooperative MIMO network, the diversity gains of the system will be sacrificed. A central node which controls the transmission power for each link is employed in \cite{O. Seong-Jun}. Although the central control power allocation can improve the performance significantly, the complexity of the calculation increases with the size of the system. The works on the power allocation problem for the DF strategy measuring the outage probability in each relay node with a single antenna and determining the power for each link between the relay nodes and the destination node, have been reported in \cite{Min Chen}-\cite{Yindi Jing}. The diversity gain can be improved by using the relay nodes with multiple antennas. When the number of relay node is the same, the cooperative gain can be improved by using the DF strategy compared with a system employing the AF strategy. However, the interference at the destination will be increased if the relay nodes forward the incorrectly detected symbols in the DF strategy.

In this paper, we propose a joint MBER adaptive power allocation algorithm with linear receiver design for cooperative MIMO systems. There are multiple relay nodes with multiple antennas to achieve an AF cooperating strategy with DSTC between the source node and the destination node. By using adaptive algorithms with the MBER criterion \cite{R. C. de Lamare2003} \cite{Chen S.}, the power allocation parameters and the linear receive filter parameters can be determined and fed back to each transmission node through a feedback channel that is assumed error free and delay free. The performance indicates the advantages in diversity gain of the proposed JPA algorithm.

The paper is organized as follows. Section II provides the two-hop cooperative MIMO system with multiple relays applying the AF strategy and DSTC scheme. Section III describes the constrained power allocation problem and the linear MBER detection method, and in Section IV, the proposed iterative SG algorithm is derived. Section V focuses on the results of the simulations and Section VI contains the conclusion.
\begin{figure}\label{Fig.1}
  \includegraphics[width=3.5in]{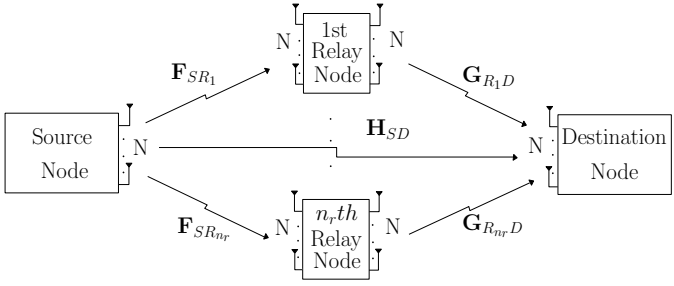}\\
  \caption{Cooperative MIMO System Model with $n_r$ Relay nodes}\label{1}
\end{figure}

\section{Cooperative System Model}

The communication system under consideration, shown in Fig.1, is a MIMO communication system transmitting through a MIMO channel from the source node to the destination node. The BPSK modulation scheme is employed in our system by mapping  the information bits ${\boldsymbol b}[i]$ to the transmitted symbols ${\boldsymbol s}[i]\in\{{\pm 1}\}$. There are $n_r$  relay nodes with $N$ antennas for transmitting and receiving, applying an AF cooperative strategy as well as a DSTC scheme, between the source node and the destination node. A two-hop communication system that transmits symbols from the source to $n_r$ relay nodes as well as to the destination node in the $n_r+1$ phases, followed by transmitting the amplified and re-encoded symbols from each relay node to the destination node in the next $n_r$ phases. We consider only one user at the source node in our system that has $N$ Spatial Multiplexing (SM)-organized data symbols contained in each packet. The received symbols at the $k-th$ relay node and the destination node are denoted as ${\boldsymbol r}_{{SR}_{k}}$ and ${\boldsymbol r}_{SD}$, respectively, where $k=1,2,...,n_r$. The received symbols ${\boldsymbol r}_{{SR}_{k}}$ will be amplified before mapped into an STC matrix. We assume that the synchronization in each node is perfect. The received symbols at the destination node and each relay node can be described as follows
\begin{equation}\label{2.1}
\begin{aligned}
{\boldsymbol r}_S[i]  = & \rm{diag}\left[\begin{array}{c} {\boldsymbol H}_{SD}[i] \\{\boldsymbol F}_{SR_1}[i] \\ \vdots \\ {\boldsymbol F}_{SR_{n_r}}[i] \end{array} \right] \left[\begin{array}{c} {\boldsymbol A}_{SD}[i] \\ {\boldsymbol A}_{SR_1}[i] \\ \vdots  \\ {\boldsymbol A}_{SR_{n_r}}[i] \end{array} \right]\left[\begin{array}{c}s_1 \\ s_2 \\ \vdots \\ s_N\end{array} \right] \\
= & [{\boldsymbol r}_{SD}[i],{\boldsymbol r}_{SR_1}[i],...,{\boldsymbol r}_{SR_{n_r}}[i]]^T \\
= & {\boldsymbol D}_S[i]{\boldsymbol A}_S[i]{\boldsymbol s}[i] + {\boldsymbol n}_S[i],
\end{aligned}
\end{equation}
\begin{equation*}
i = 1,2,~...~,N,
\end{equation*}
where the $\rm{diag}(\cdot)$ operator transforms the argument into a diagonal structure and the $(n_r+1)N \times 1$ vector ${\boldsymbol n}_S[i]$ denotes the zero mean complex circular symmetric Additive White Gaussian Noise (AWGN) vector with variance $\sigma^{2}$. The transmitted symbol vector ${\boldsymbol s}[i]$ contains $N$ parameters, ${\boldsymbol s}[i] = [s_{1}[i], s_{2}[i], ... , s_{N}[i]]^T$, which have a covariance matrix $E\big[ {\boldsymbol s}[i]{\boldsymbol s}^{H}[i]\big] = \sigma_{s}^{2}{\boldsymbol I}_N$, where $E[\cdot]$ stands for expected value, $(\cdot)^H$ denotes the Hermitian operator, $\sigma_s^2$ is the signal power which we assume to be equal to 1 and ${\boldsymbol I}_N$ is the $N \times N$ identity matrix. The diagonal matrix ${\boldsymbol D}_{S}[i]$ is the $(n_r+1)N \times n_rN$ channel gain matrix that contains the $N \times N$ channel matrix ${\boldsymbol H}[i]$ between the source node and the destination node and the channel matrix ${\boldsymbol F}_k[i]$ between the source node and each relay node, where $k=1,2,...,n_r$. The $n_rN \times N$ matrix ${\boldsymbol A}_{S}[i]$ denotes the power allocation matrix for the transmitted symbol vector ${\boldsymbol s}[i]$. Notice that the $(n_r+1)N \times 1$ vector ${\boldsymbol r}_S[i]$ in (\ref{2.1}) contains the received symbol vector at the destination node and the relay nodes, which is ${\boldsymbol r}_S[i]=[{\boldsymbol r}_{SD}[i], {\boldsymbol r}_{SR_1}[i], ... , {\boldsymbol r}_{SR_{n_r}}[i]]^T$.

After processing the received vector ${\boldsymbol r}_{SR_k}[i]$ at the $k-th$ relay node, the signal vector $\tilde{{\boldsymbol s}}_{SR_k}[i]$ can be obtained and will be forwarded to the destination node. A power allocation parameter vector will be assigned to $\tilde{{\boldsymbol s}}_{SR_k}[i]$ first, and then the amplified symbols in $\tilde{{\boldsymbol s}}_{SR_k}[i]$ will be re-encoded by a $N \times T$ DSTC scheme ${\boldsymbol M}(\tilde{{\boldsymbol s}})$ and then forwarded to the destination node. The relationship between the $k-th$ relay and the destination node can be described as
\begin{equation}\label{2.2}
{\boldsymbol R}_{R_{k}D}[i] = {\boldsymbol G}_k[i]{\boldsymbol M}_{R_{k}D}[i] + {\boldsymbol N}_{R_{k}D}[i],
\end{equation}
\begin{equation*}
    k = 1,2, ..., n_r
\end{equation*}
where the $N \times T$ matrix ${\boldsymbol M}_{R_{k}D}[i]$ is the DSTC matrix employed in the relays whose elements are the amplified symbols in $\tilde{{\boldsymbol s}}_{SR_k}[i]$. The $N \times T$ received symbol matrix ${\boldsymbol R}_{R_{k}D}[i]$ in (\ref{2.2}) can be transformed and denoted as a $NT \times 1$ vector ${\boldsymbol r}_{R_{k}D}[i]$ given by
\begin{equation}\label{2.3}
{\boldsymbol r}_{R_{k}D}[i]  = {\boldsymbol G}'_k[i]{\boldsymbol A}_{R_{k}D}[i]\tilde{{\boldsymbol s}}_{SR_k}[i] + {\boldsymbol n}_{R_{k}D}[i],
\end{equation}
\begin{equation*}
 k = 1,2,...,n_r
\end{equation*}
where the diagonal $N \times N$ matrix ${\boldsymbol A}_{R_{k}D}[i] = {\rm diag} [\alpha_{1_{R_{k}D}}[i], ... , \alpha_{N_{R_{k}D}}[i]]$ contains the power allocation parameters assigned for the $j-th$ symbol $\tilde{s}_{j_{{SR_k}}}[i]$. The $NT \times N$ matrix ${\boldsymbol G}'_k[i]$ stands for the equivalent channel matrix which is the DSTC scheme ${\boldsymbol M}(\tilde{{\boldsymbol s}}[i])$ combined with the channel matrix ${\boldsymbol G}_{R_{k}D}[i]$. The $NT \times 1$ noise vector ${\boldsymbol n}_{R_{k}D}[i]$ is an equivalent noise vector at the destination node which contains the noise parameters in ${\boldsymbol N}_{R_{k}D}[i]$. After rewriting ${\boldsymbol R}_{R_{k}D}[i]$ we can consider the received symbol vector at the destination node as $n_r+1$ parts, one is from the source node and the remaining $n_r$ are from the relay nodes, and write the received symbol vector for cooperative detection as
\begin{equation}
\begin{aligned}
{\boldsymbol r}[i]  &=
\left[\begin{array}{c} {\boldsymbol H}[i]{\boldsymbol A}_{SD}[i]{\boldsymbol s}[i]  \\ {\boldsymbol G}'_1[i]{\boldsymbol A}_{R_{1}D}[i]\tilde{{\boldsymbol s}}_{SR_1}[i]  \\ {\boldsymbol G}'_2[i]{\boldsymbol A}_{R_{2}D}[i]\tilde{{\boldsymbol s}}_{SR_2}[i] \\ \vdots \\ {\boldsymbol G}'_{n_r}[i]{\boldsymbol A}_{R_{n_r}D}[i]\tilde{{\boldsymbol s}}_{SR_{n_r}}[i] \end{array} \right] + \left[\begin{array}{c}{\boldsymbol n}_{SD}[i] \\ {\boldsymbol n}_{R_{1}D}[i] \\ {\boldsymbol n}_{R_{2}D}[i] \\ \vdots \\ {\boldsymbol n}_{R_{n_r}D}[i] \end{array} \right] \\
& = {\boldsymbol D}_D[i]{\boldsymbol A}_D[i]{\boldsymbol s}_D[i] + {\boldsymbol n}_D[i],
\end{aligned}
\end{equation}
where the $(n_rT + 1)N \times (n_r + 1)N$ diagonal matrix ${\boldsymbol D}_D[i]$ contains the channel gain of all the links between the relays and the destination node. The $(n_r + 1)N \times N$ power allocation parameter matrix ${\boldsymbol A}_D[i] = [{\boldsymbol A}_{SD}[i], {\boldsymbol A}_{R_{1}D}[i],~...~,~{\boldsymbol A}_{R_{n_r}D}[i]]^T$ is constructed by all the power allocation matrices between all the relay nodes and the destination node.

\section{Joint linear MBER receiver design with Power Allocation}
The MBER receiver design with power allocation for every link in all the phases between the transmitting nodes and the receiving nodes is derived as follows. If we define a $(n_rT + 1)N \times 1$ parameter vector ${\boldsymbol w}_{j}[i] = [{\boldsymbol w}_{j_1}[i],{\boldsymbol w}_{j_2}[i],...,{\boldsymbol w}_{j_{n_r+1}}[i]]$ for $i = 1, 2, ..., N$ to determine the $j-th$ symbol $s_j[i]$, by using (4), the desired information symbols at the destination node can be computed as
\begin{equation}
b_j[i] = \rm{sgn}({\boldsymbol w}^H_j[i]{\boldsymbol r}[i])=\rm{sgn}(\tilde{s}_j[i]),
\end{equation}
where $\rm{sgn}(\cdot)$ is the sign function and $\tilde{s}_j[i]$ denotes the detected symbol at the receiver which can be further written as
\begin{equation}
\begin{aligned}
    \tilde{s}_j[i] & = {\boldsymbol w}^H_j[i]{\boldsymbol r}[i] \\
    & = {\boldsymbol w}^H_j[i]({\boldsymbol D}_D[i]{\boldsymbol A}_D[i]{\boldsymbol s}_D[i] + {\boldsymbol n}_D[i]) \\
    & = {\boldsymbol w}^H_j[i]{\boldsymbol D}_D[i]{\boldsymbol A}_D[i]{\boldsymbol s}_D[i] + {\boldsymbol w}^H_j[i]{\boldsymbol n}_D[i] \\
    & = \tilde{s'}_j[i] + e_j[i],
\end{aligned}
\end{equation}
where $\tilde{s'}_j[i]$ is the noise-free detected symbol, and $e_j[i]$ denotes the error factor for the $j-th$ detected symbol $\tilde{s'}_j[i]$ with zero mean and variance $\sigma^2_n{\boldsymbol w}^H_j[i]{\boldsymbol w}_j[i]$, where $\sigma^2_n$ is the variance of the received noise ${\boldsymbol n}_D[i]$. Define a $N \times N_b$ matrix $\boldsymbol {\bar{S}}$ which is constructed by a set of vectors ${\boldsymbol {\bar{s}}}$ and $N_b=2^N$, containing all the possible combinations of the transmitted symbol vector ${\boldsymbol s}[i]$ and we can obtain
\begin{equation}
\begin{aligned}
    \bar{s}_j[i] & = {\boldsymbol w}^H_j[i]\boldsymbol {\bar{r}}[i]\\ & = {\boldsymbol w}^H_j[i]{\boldsymbol D}_D[i]{\boldsymbol A}_D[i]{\boldsymbol {\bar{s}}_l}[i],
\end{aligned}
\end{equation}
\begin{equation*}
    l = 1,2,...,N_b,
\end{equation*}
where $\bar{s}_j[i]$ is an element from the noise-free detected vector $\boldsymbol {\bar{s}}_l[i]$. To derive the BER expression for the linear filter ${\boldsymbol w}_j[i]$ and the power allocation parameter matrix ${\boldsymbol A}_D[i]$, we should determine the probability density function (pdf) of $\boldsymbol {\bar{r}}[i]$ which is
\begin{equation}\label{2.4}
\begin{aligned}
    p_{\boldsymbol {\bar{r}}[i]} = & \frac{1}{N_b\sqrt{2\pi\sigma^2_{n}{\boldsymbol w}^H_j[i]{\boldsymbol w}_j[i]}} \\
    & \sum_{l=1}^{N_b}\rm{exp}\big(-\frac{(\tilde{s}_j[i]-sgn(b_j[i])\bar{s}_j[i])^2}{2\sigma^2_{n}{\boldsymbol w}^H_j[i]{\boldsymbol w}_j[i]}\big).
\end{aligned}
\end{equation}
As a result we can obtain the bit error rate (BER) expression as
\begin{equation}\label{2.5}
    P_E({\boldsymbol w}_j[i],\alpha_j[i])=\frac{1}{N_b}\sum^{N_b}_{l=1}{\boldsymbol Q}(c_j[i]),
\end{equation}
where
\begin{equation}\label{2.6}
    c_j[i] = \frac{sgn(s_j[i])\bar{s}_j[i]}{\sigma_{n}\sqrt{{\boldsymbol w}^H_j[i]{\boldsymbol w}_j[i]}}.
\end{equation}
Before we derive the adaptive algorithm, we have to specify the power allocation parameter $\alpha_j[i]$ in (\ref{2.5}). The power allocation matrices, ${\boldsymbol A}_{SD}[i]$, ${\boldsymbol A}_{SR_k}[i]$ and ${\boldsymbol A}_{R_kD}[i]$, where $k=1,2,...,n_r$, have to be estimated in our algorithm. For simplicity of expression, we use the factor $\alpha$ in our estimation function instead of the power allocation matrices. It is shown that in what follows, the parameters $\alpha_{j_{SD}}[i]$, $\alpha_{j_{SR_k}}[i]$ and $\alpha_{j_{R_kD}}[i]$ denote the power allocation parameters assigned for the $j-th$ signal in ${\boldsymbol A}_{SD}[i]$, ${\boldsymbol A}_{SR_k}[i]$ and ${\boldsymbol A}_{R_kD}[i]$, as described by
\begin{equation*}\label{a1}
\begin{aligned}
    & {\boldsymbol A}_{SD}[i]=\rm{diag} \left[\begin{array}{c}\alpha_{1_{SD}}[i]\\ \alpha_{2_{SD}}[i] \\  \vdots \\ \alpha_{{n_r}_{SD}}[i] \end{array} \right],
    {\boldsymbol A}_{SR_k}[i]=\rm{diag}\left[\begin{array}{c}\alpha_{1_{SR_k}}[i]\\ \alpha_{2_{SR_k}}[i] \\  \vdots \\ \alpha_{{n_r}_{SR_k}}[i] \end{array} \right], \\
    & {\boldsymbol A}_{R_kD}[i]=\rm{diag} \left[\begin{array}{c}\alpha_{1_{R_kD}}[i]\\ \alpha_{2_{R_kD}}[i] \\  \vdots \\ \alpha_{{n_r}_{R_kD}}[i] \end{array} \right].
\end{aligned}
\end{equation*}
By substituting (\ref{2.4}) and (\ref{2.6}) into (\ref{2.5}) and taking the gradient with respect to different arguments, we can obtain (\ref{t1}) - (\ref{t4}),
\begin{table*}
\begin{equation}\label{t1}
    \nabla P_{E_{{\boldsymbol w}_j}}[i]  = \frac{1}{N_b\sqrt {2\pi}}\sum^{N_b}_{l=1}exp\big(-\frac{c^2_j[i]}{2}\big)
    \frac{sgn(s_j[i])\boldsymbol {\bar{r}}[i]\sigma_{n} \sqrt{{\boldsymbol w}^H_j[i]{\boldsymbol w}_j[i]} - sgn(s_j[i]){\boldsymbol w}^H_j[i]\boldsymbol {\bar{r}}[i]\sigma_{n}({\boldsymbol w}^H_j[i]{\boldsymbol w}_j[i])^{-\frac{1}{2}}{\boldsymbol w}_j[i]}{\sigma^2_{n}{\boldsymbol w}^H_j[i]{\boldsymbol w}_j[i]}
\end{equation}
\begin{equation}\label{t2}
    \nabla P_{E_{{\alpha}_{j_{SD}}}}[i] = \frac{1}{\sqrt {2\pi}}\sum^{N_b}_{l=1}exp\big(-\frac{c^2_j[i]}{2}\big)\frac{sgn(s_j[i]){\boldsymbol w}^H_{j_1}{\boldsymbol h}_j[i]\bar{s}_j[i]}{\sigma_{n}\sqrt{{\boldsymbol w}^H_j[i]{\boldsymbol w}_j[i]}}
\end{equation}
\begin{equation}\label{t3}
    \nabla P_{E_{{\alpha}_{j_{SR_k}}}}[i] = \frac{1}{\sqrt {2\pi}}\sum^{N_b}_{l=1}exp\big(-\frac{c^2_j[i]}{2}\big)\frac{sgn(s_j[i]){\boldsymbol w}^H_{j_k}{\boldsymbol g}_{{R_kD}_j}[i]{\alpha}_{j_{R_kD}}[i]{\boldsymbol f}_{{SR_k}_j}[i]\bar{s}_j[i]}{\sigma_{n}\sqrt{{\boldsymbol w}^H_j[i]{\boldsymbol w}_j[i]}}
\end{equation}
\begin{equation}\label{t4}
    \nabla P_{E_{{\alpha}_{j_{R_kD}}}}[i] = \frac{1}{\sqrt {2\pi}}\sum^{N_b}_{l=1}exp\big(-\frac{c^2_j[i]}{2}\big)\frac{sgn(s_j[i]){\boldsymbol w}^H_{j_k}{\boldsymbol d}_{{k_D}_j}[i]{\alpha}_{j_{SR_k}}[i]\bar{s}_j[i]}{\sigma_{n}\sqrt{{\boldsymbol w}^H_j[i]{\boldsymbol w}_j[i]}}
\end{equation}
\rule{18cm}{1pt}
\end{table*}
where ${\boldsymbol h}_j[i]$ is the $j-th$ column of ${\boldsymbol H}[i]$ and ${\boldsymbol g}_{{R_kD}_j}[i]$ and ${\boldsymbol f}_{{SR_k}_j}[i]$ denote the $j-th$ column of ${\boldsymbol G}'_{R_kD}[i]$ and ${\boldsymbol F}_{SR_k}[i]$, respectively. By making use of (\ref{t1}) - (\ref{t4}) and an SG algorithm in \cite{S. Haykin}, the updated receive filter ${\boldsymbol w}_j[i]$ and the power allocation parameters assigned for the $j-th$ transmitted symbol$s_j[i]$ between the source node and the destination node ${\alpha}_{j_{SD}}[i]$, between the source node and the $k-th$ relay node ${\alpha}_{j_{SR_k}}[i]$, and between the $k-th$ relay node and the destination node ${\alpha}_{j_{R_kD}}[i]$, can be obtained as follows
\begin{equation}\label{a-1}
\begin{aligned}
{\boldsymbol w}_j[i + 1] & = {\boldsymbol w}_j[i] - \mu\nabla P_{E_{{\boldsymbol w}_j}}[i], \\
\end{aligned}
\end{equation}
\begin{equation}\label{a-2}
\begin{aligned}
\alpha_{j_{SD}}[i + 1] & = \alpha_{j_{SD}}[i] - \gamma \nabla P_{E_{{\alpha}_{j_{SD}}}}[i], \\
\end{aligned}
\end{equation}
\begin{equation}\label{a-3}
\begin{aligned}
\alpha_{j_{R_{k}D}}[i + 1] & = \alpha_{j_{R_kD}}[i] - \gamma \nabla P_{E_{{\alpha}_{j_{R_kD}}}}[i], \\
\end{aligned}
\end{equation}
\begin{equation}\label{a-4}
\begin{aligned}
\alpha_{j_{SR_k}}[i + 1] & = \alpha_{j_{SR_k}}[i] - \gamma \nabla P_{E_{{\alpha}_{j_{SR_k}}}}[i], \\
\end{aligned}
\end{equation}
where $\mu$ and $\gamma$ are step sizes for estimating the receive filter parameters and the power allocation parameters, respectively.

\subsection{Adaptive MBER SG Estimation and Power Allocation}
The key to achieving the adaptive SG estimation algorithm described in (\ref{a-1})-(\ref{a-4}) is to find out an efficient and reliable method to calculate the pdf of the received symbol vector ${\boldsymbol r}[i]$ at the destination node. In \cite{Parzen}-\cite{Bowman}, the Parzen window and kernel density estimation method are introduced, which can guarantee the accuracy of the probability distribution. Because the noise vector in (4) is Gaussian and the kernel density estimation can provide the reliable pdf estimation when dealing with Gaussian mixtures, we choose the Kernel density estimation method in our study.

By transmitting a block of $K$ training samples ${\boldsymbol {\hat{s}}}[i]=\rm{sgn}({\boldsymbol {\hat{b}}})$, the kernel density estimated pdf of ${\boldsymbol {\hat{s}}}[i]$ is given by
\begin{equation}
\begin{aligned}
    p_{\hat{s}[i]} = & \frac{1}{K\sqrt{2\pi}\rho_n\sqrt{{\boldsymbol w}^H_j[i]{\boldsymbol w}_j[i]}} \\
    & \sum_{l=1}^{K}exp\big(-\frac{(\tilde{s}_j[i]-sgn(b_j[i])\hat{s}_j[i])^2}{2\rho^2_n{\boldsymbol w}^H_j[i]{\boldsymbol w}_j[i]}\big),
\end{aligned}
\end{equation}
where $\rho_n$ is related to the standard deviation of noise $\sigma_n$ and it is suggested in \cite{Parzen} that a lower bound of $\rho_n=\big(\frac{4}{3K}\big)^{\frac{1}{5}}\sigma_n$ should be chosen. The expression of the BER can be derived as
\begin{equation}
    \hat{P_E}({\boldsymbol w}_j[i],\alpha_j[i])=\frac{1}{K}\sum^{K}_{l=1}{\boldsymbol Q}(c_{k_j}[i]),
\end{equation}
where
\begin{equation}
    c_{k_j}[i] = \frac{sgn(\hat{s}_j[i])\bar{\hat{s}}_j[i]}{\rho_n\sqrt{{\boldsymbol w}^H_j[i]{\boldsymbol w}_j[i]}}.
\end{equation}
Then by taking the gradient of (20) with respect to different arguments we can obtain the gradient value of the estimated error probability, and by iterative calculation of the parameters in the receive filter vector with the power allocation parameters, we can finally achieve the optimal joint estimation result.

\subsection{Adaptive SG Channel Estimation}
In this subsection we will derive an adaptive SG algorithm for estimating the equivalent channel matrix ${\boldsymbol D}_D[i]$. The channel estimation can be described as an optimization problem
\begin{equation}\label{c1}
{\boldsymbol D}_D[i] = \arg \min_{{\boldsymbol D}_D[i]}{E\big [||{\boldsymbol r}[i] - {\boldsymbol D}_D[i]{\boldsymbol A}[i]_D{\boldsymbol s}_D[i]||^2]}.
\end{equation}
Define the received symbol vector ${\boldsymbol r}[i] = [{\boldsymbol r}_1[i],...,{\boldsymbol r}_{n_r+1}[i]]$. The optimization problem described in (\ref{c1}) can be divided into three parts, which correspond to individually computing ${\boldsymbol H}[i]$ and ${\boldsymbol G}'_{R_kD}[i]$. We then derive an SG algorithm by calculating the cost function ${\rm C}_{\boldsymbol H}$ and ${\rm C}_{{\boldsymbol G}'_k}$
\begin{equation}
 C_{\boldsymbol H}  =  E\big [||{\boldsymbol r}_1[i] - {\boldsymbol H}[i]{\boldsymbol A}_{SD}[i]{\boldsymbol s}[i]||^2\big ],
\end{equation}
\begin{equation}
\begin{aligned}
 C_{{\boldsymbol G}'_{k}} = & E\big [||{\boldsymbol r}_k[i] - {\boldsymbol G}'_k[i]{\boldsymbol A}_{R_{k}D}[i]{\boldsymbol s}[i]||^2\big ].
 \end{aligned}
\end{equation}
Then by taking instantaneous gradient terms of ${\rm C}_{\boldsymbol H}$ and ${\rm C}_{{\boldsymbol G}'_k}$ with respect to ${\boldsymbol H}[i]$ and the $j-th$ column of the equivalent channel matrix ${\boldsymbol G}'_{R_kD}[i]$, we can obtain
\begin{equation}
\nabla C_{{\boldsymbol H}^*[i]} = -{\boldsymbol A}_{SD}[i]{\boldsymbol s}[i]({\boldsymbol r}_1[i] - {\boldsymbol H}[i]{\boldsymbol A}_{SD}[i]{\boldsymbol s}[i]),
\end{equation}
\begin{equation}
\nabla C_{{{\boldsymbol G}'}^*_k[i]} = - {\boldsymbol A}_{k_D}[i]{\boldsymbol s}[i]({\boldsymbol r}_k[i] - {\boldsymbol G}'_{R_kD}[i]{\boldsymbol A}_{R_kD}[i]{\boldsymbol s}[i]).
\end{equation}
Considering the SG descent rules \cite{S. Haykin} and the results of gradient terms we can obtain the adaptive SG channel estimation expressions which are
\begin{equation}
{\boldsymbol H}[i + 1] = {\boldsymbol H}[i] - \beta\nabla C_{{\boldsymbol H}^*[i]}
\end{equation}
\begin{equation}
    {{\boldsymbol G}'}_k[i + 1] = {\boldsymbol G}'_k[i] - \beta\nabla C_{{{\boldsymbol G}'}^*_k[i]}
\end{equation}
where $\beta$ is the step size of the recursion. The adaptive SG algorithm for the equivalent channel ${\boldsymbol H}[i]$ and ${\boldsymbol G}'_k[i]$ requires the calculation complexity of $({\boldsymbol O}(N^2))$ and $({\boldsymbol O}(TN))$.

\section{Simulations}

The simulation results are provided in this section to assess the
proposed algorithm. The BER performance of the adaptive JPA and the
EPA using a linear MBER receive filter (JPA-MBER) and (EPA-MBER)
algorithms with the power constraint employs different numbers of
relay nodes and different DSTC schemes in \cite{Birsen
Sirkeci-Mergen} are compared with $n_r = 1,2$ relay nodes in the
simulation. In the simulation we define the power constraint $P_T$
as equal to 1, and the number of antennas $N=2$ at each node. In
Fig. 2, the Almouti STBC scheme is used at relay nodes in the
simulation. The results illustrate that the performance of the
EPA-MBER and JPA-MBER algorithms is superior to that of MMSE, and
the JPA-MBER algorithm outperforms the EPA-MBER algorithm using the
same DSTC scheme to achieve the identical BER. There is about 2.5 dB
of gain compared the JPA-MBER algorithm with the EPA-MBER one when
using 1 relay node to achieve the BER in $10^{-3}$. The performance
improvement of the proposed JPA-MBER algorithm is achieved with more
relays employed in the system as an increased spatial diversity is
provided by the relays. The coding gain achieved by using different
DSTC encoders at the relay nodes, namely, the Alamouti and the
randomized Alamouti (R-Alamouti) is shown in the results of Fig. 3.
The number of relay nodes is equal to 2. The JPA-MBER algorithm
outperforms the EPA-MBER algorithm when using a different STC
scheme, and the JPA-MBER algorithm with R-Alamouti STBC scheme
obtains 1 dB of gain compared to the that of the system with the
Alamouti STBC scheme. The proposed adaptive algorithm can be used in
the cooperative MIMO system with linear detector in
\cite{delamare_spadf} or non-linear detector in  \cite{vikalo}. With
different criteria in \cite{delamarespl07} to \cite{Tong P}, the
proposed adaptive JPA algorithm can obtain more diversity gains
compared with the EPA algorithms.

\section{Conclusion}

We have proposed a joint power allocation and receiver design algorithm using a linear MBER receive filter with the power constraint between the source node and the relay nodes, and between relay nodes and the destination node. A joint iterative estimation algorithm for computing the power allocation parameters and linear MBER receive filter vector has been derived. The simulation results illustrate the advantage of the proposed power allocation algorithm by comparing it with the equal power algorithm. The proposed algorithm can be utilized with different DSTC schemes and a variety of detectors \cite{vikalo} \cite{delamare_spadf} and estimation algorithms \cite{delamarespl07} \cite{jidf} in the cooperative MIMO systems with AF strategy and can also be extended to the DF cooperation protocols.



\ifCLASSOPTIONcaptionsoff
  \newpage
\fi

\end{document}